\documentclass[a4paper,10pt,prl,twocolumn,showpacs,aps,floats,floatfix,superscriptaddress]{revtex4}

\usepackage{epsfig}
\begin{document}

\newcommand{\avk}{\langle k \rangle}
\newcommand{\fluck}{\langle k^2 \rangle}

\title{Velocity and hierarchical spread of epidemic outbreaks
in scale-free networks}

\author{Marc Barth{\'e}lemy} \affiliation{CEA-Centre d'Etudes de
  Bruy{\`e}res-le-Ch{\^a}tel, D{\'e}partement de Physique Th{\'e}orique et
  Appliqu{\'e}e BP12, 91680 Bruy{\`e}res-Le-Ch{\^a}tel, France} \author{Alain
  Barrat} \affiliation{Laboratoire de Physique Th{\'e}orique (UMR du CNRS
  8627), Batiment 210, Universit{\'e} de Paris-Sud 91405 Orsay, France}
\author{Romualdo Pastor-Satorras} \affiliation{Departament de F\'isica i
  Enginyeria Nuclear Universitat Polit{\`e}cnica de Catalunya, Campus Nord
  08034 Barcelona, Spain} \author{Alessandro Vespignani}
\affiliation{Laboratoire de Physique Th{\'e}orique (UMR du CNRS 8627),
  Batiment 210, Universit{\'e} de Paris-Sud 91405 Orsay, France}


\widetext
\begin{abstract}
  We study the effect of the connectivity pattern of complex networks
  on the propagation dynamics of epidemics.  The growth time scale of
  outbreaks is inversely proportional to the network degree
  fluctuations, signaling that epidemics spread almost instantaneously
  in networks with scale-free degree distributions. This feature is
  associated to an epidemic propagation that follows a precise
  hierarchical dynamics. Once the highly connected hubs are reached,
  the infection pervades the network in a progressive cascade across
  smaller degree classes.  The present results are relevant for the
  development of adaptive containment strategies.
\end{abstract}


\maketitle 

The connectivity pattern of the network of individual contacts has
been acknowledged since long as a relevant factor in determining the
properties of epidemic spreading
phenomena~\cite{Hethcote:1984,May:1992}.  This issue assumes a
particular relevance in the case of networks characterized by complex
topologies and very heterogeneous
structures~\cite{barabasi:2000,mdbook,psvbook,lil,may01} that in many
cases present us with new epidemic propagation
scenarios~\cite{Moore:2000,Abramson:2001,Pastor:2001,lloyd01,Moreno:2002,newman02}.
A striking example of this situation is provided by scale-free
networks characterized by large fluctuations in the number of
connections (degree) $k$ of each vertex. This feature usually finds
its signature in a heavy-tailed degree distribution with power-law
behavior of the form $P(k)\sim k^{-\gamma}$, with $2\leq\gamma\leq 3$,
that implies a non-vanishing probability of finding vertices with very
large degrees~\cite{barabasi:2000,mdbook,psvbook,Amaral:2000}.  The
latter are the ``hubs'' or ``superspreaders''~\cite{Hethcote:1984}
responsible for the proliferation of infected individuals whatever the
rate of infection characterizing the epidemic, eventually leading to
the absence of any epidemic threshold below which the infection cannot
initiate a major outbreak \cite{Pastor:2001}.  This new scenario is of
practical interest in computer virus diffusion and the spreading of
diseases in heterogeneous populations~\cite{mdbook,psvbook,lil,may01}.
It also raises new questions on how to protect the network and find
optimal strategies for the deployment of immunization
resources~\cite{Pastor:2002,dezso02,havlin03}.  So far, however,
studies of epidemic models in complex networks have been focused on
the stationary properties of endemic states or the final prevalence
(number of infected individuals) of epidemics.  The dynamical
evolution of the outbreaks has been instead far less investigated and
a detailed inspection of the temporal behavior in complex topologies
is still missing.

In this paper we intend to fill this gap by providing a first analysis
of the time evolution of epidemic outbreaks in complex networks with 
highly heterogeneous connectivity patterns. We consider the time
behavior of
epidemic outbreaks and find that the growth of infected individuals is
governed by a time scale $\tau$ proportional to the ratio between the
first and second moment of the network's degree distribution, $\tau\sim
\avk/\fluck$.  This implies an instantaneous rise of the
prevalence in very heterogeneous networks where $\fluck\to\infty$ in
the infinite size limit. In particular, this result shows that 
scale-free networks with $2\leq\gamma\leq 3$ exhibit, 
along with the lack of an intrinsic epidemic 
threshold, a virtually infinite propagation velocity of the infection.  
Furthermore, we study the
detailed propagation in time of the infection through the different
degree classes in the population. We find a striking hierarchical
dynamics in which the infection propagates via a cascade that
progresses from higher to lower degree classes. This infection
hierarchy might be used to develop dynamical ad-hoc strategies for
network protection.

In order to study the dynamical evolution of epidemic outbreaks we shall
focus on the susceptible-infected (SI) model in which individuals can be in
two discrete states, either susceptible or
infected~\cite{May:1992,Murray:1993}.  Each individual is represented by a
vertex of the network and the edges are the connections between individuals
along which the infection may spread. The total population 
(the size of the network) $N$ is assumed to
be constant and if $S(t)$ and $I(t)$ are the number of susceptible and
infected individuals at time $t$, respectively, then $N=S(t)+I(t)$. In the SI
model the infection transmission is defined by the spreading rate $\lambda$
at which susceptible individuals acquire the infection from an infected
neighbor.  In this model infected individuals remain always infective, an
approximation that is useful to describe early epidemic stages in which no
control measures are deployed. We shall see in the following that the results
obtained for the SI model can be readily generalized to more complex schemes.

A first analytical description of the SI model can be undertaken
within the homogeneous mixing hypothesis~\cite{May:1992,Murray:1993},
consisting in a mean-field description of the system in which all
vertices are considered as being equivalent. In this case the reaction rate
equation for the average density of infected individuals $i(t)=I(t)/N$ (the
prevalence) reads as
\begin{equation}
  \frac{d i(t)}{dt}=\lambda \avk i(t)\left[1-i(t)\right].
  \label{SI-MF}
\end{equation}
The above equation states that the growth rate of infected individuals
is proportional to the spreading rate $\lambda$, the density of susceptible
vertices that may become infected, $s(t)=1-i(t)$, and the number of
infected individuals in contact with any susceptible vertex.  The
homogeneous mixing hypothesis considers that this last term is simply
the product of the number of neighbors $\avk$ and the average density
$i(t)$.  Obviously, this approximation neglects correlations among
individuals and considers that all vertices have the same number of
neighbors $\avk$, i.e. it assumes a perfectly homogeneous network.  At
small times, when the density of infected vertices is very small, we
can neglect terms of order ${\cal O}(i^2)$ and obtain the leading
behavior $i(t)\simeq i_0 e^{t/\tau_{H}}$, where $i_0$ is the initial density
of infected individuals and $\tau_H= (\lambda \avk)^{-1}$ is the time
scale governing the growth of the infection in a homogeneous network.
\begin{figure}[t]
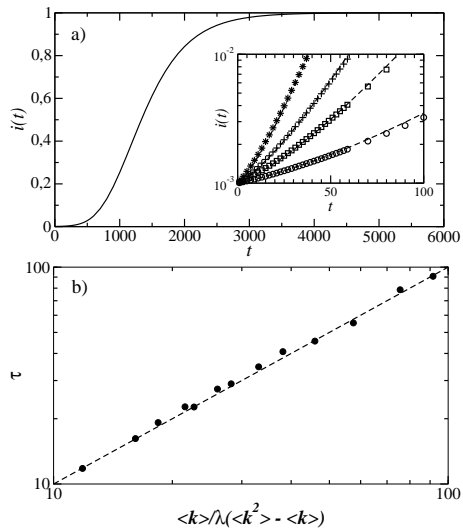

\begin{center}
\epsfig{file=fig1a.eps,width=6cm}
\epsfig{file=fig1b.eps,width=6cm}
\end{center}
\caption{a) Average density of infected individuals  
  versus time in a BA network of N=$10^4$ with $m=2$.  The inset shows
  the exponential fit obtained in the early times (lines) and the
  numerical curves $i(t)$ for networks with m=4,8,12,20 (from bottom
  to top).  b) Measured time scale $\tau$ in BA networks as obtained
  from exponential fitting versus the theoretical prediction for
  different values of $m$ and $N$ corresponding to different levels of
  heterogeneity.}
\label{fig1}
\end{figure}

The above calculations recover that the outbreak's time scale $\tau_H$
is inversely proportional to the epidemic reproduction rate; i.e. the
mean number of infections generated by each infected individual.
However, in populations with a heterogeneous connectivity pattern, it
is known that the reproduction rate depends upon the contacts'
fluctuations~\cite{May:1992}, and it is natural to expect that also
the outbreaks' time scale is analogously affected.  Indeed, in
heterogeneous networks the degree $k$ of vertices is highly
fluctuating and the average degree is not anymore a meaningful
characterization of the network properties.  In order to take fully
into account the degree heterogeneity in the dynamical evolution, it
is possible to write down the whole set of reaction rate equations for
the average densities of infected vertices of degree $k$, $i_k(t)=I_k(t)/N_k$,
where $N_k$ and $I_k(t)$ are the number of vertices and infected
vertices within each degree class $k$,
respectively~\cite{Pastor:2001}. For the SI model the evolution
equations read as
\begin{equation}
  \frac{d i_k(t)}{dt}=\lambda k \left[1-i_k(t)\right]\Theta_k(t),
  \label{SIk}
\end{equation}
where the creation term is proportional to the spreading rate
$\lambda$, the degree $k$, the probability $1-i_k$ that a vertex with
degree $k$ is not infected, and the density $\Theta_k$ of infected
neighbors of a vertex of degree $k$. The latter term is thus the
average probability that any given neighbor of a vertex of degree $k$
is infected.  In the case of uncorrelated networks~\cite{corr},
$\Theta_k \equiv \Theta$ is independent of the degree of the emanating
edge. The probability that each edge of a susceptible is pointing to
an infected vertex of degree $k'$ is proportional to the fraction of
edges emanated from these vertices.  By considering that at least one
of the edge of each infected vertex is pointing to another infected
vertex, from which the infection has been transmitted, one obtains
\begin{equation}
  \Theta(t)=\frac{\sum_{k'} (k'-1)P(k') i_{k'}(t)}{\avk},
\label{theta}
\end{equation}
where $\langle k \rangle =\sum_{k'} k'P(k')$ is the proper normalization 
factor dictated by the total number of edges.
A reaction rate equation for $\Theta(t)$ can be obtained from
Eqs.~(\ref{SIk}) and~(\ref{theta}). In the initial epidemic stages,
we neglect terms of order ${\cal O}(i^2)$, obtaining the following set of
equations 
\begin{eqnarray}
\frac{d i_k(t)}{dt}&=&\lambda k \Theta(t),\\
\frac{d \Theta(t)}{dt}&=&\lambda \left(\frac{\fluck}{\avk}-1\right)\Theta(t).
\end{eqnarray}
These equations can be solved with the uniform initial condition
$i_k(t=0) =i_0$ yielding  for the total
average prevalence $i(t) = \sum_k P(k) i_k(t)$ 
\begin{equation}
  i(t) =i_0 \left[1+\frac{\avk^2-\avk}{\fluck-\avk}(e^{t/\tau}-1)\right],
\end{equation}
where 
\begin{equation}
  \tau = \frac{\avk}{\lambda(\fluck-\avk)}.
\label{tauSF}
\end{equation}
This readily implies that the growth time scale of an epidemic
outbreak is related to the graph heterogeneity. Indeed, the ratio
$\kappa=\fluck / \avk$ is the parameter defining the level of
heterogeneity of the network, since the normalized degree variance can
be expressed as $\kappa/\avk-1$ and therefore high levels of fluctuations
correspond to $\kappa\gg\avk$. In  networks with a Poisson degree
distribution in which $\kappa=\avk +1$, we recover the result $\tau \simeq
(\lambda \avk)^{-1}$.  Instead, in networks with very heterogeneous
connectivity patterns $\kappa$ is very large and the outbreak time-scale
$\tau$ is very small, signaling a very fast diffusion of the infection.
In particular, in scale-free networks characterized by a degree
exponent $2\leq\gamma\leq 3$ we have that 
$\kappa\sim\fluck\to\infty$ with the network
size $N\to\infty$.  Therefore, while $\tau$ is a function of the finite size
$N$, for large scale-free networks we face a virtually instantaneous
rise of the epidemic incidence.

It is worth stressing that these results can be easily extended to the
susceptible-infected-susceptible (SIS) and the
susceptible-infected-removed models (SIR) \cite{May:1992}, in which 
the Eq.~(\ref{SIk}) contains an extra term $-\mu
i_k(t)$ defining the rate at which infected individuals of degree $k$
recover and become again susceptible or removed from the 
population, respectively. In addition, in the SIR
model the normalization imposes that $s_k(t)=1-i_k(t)-r_k(t)$, where
$r_k(t)$ is the density of removed individuals of degree $k$.  The
inclusion of the decaying term $-\mu i_k$, does not change the picture
obtained in the SI model.  By using the same approximations, the time
scale is found to behave as $\tau \sim
\avk/(\lambda\fluck-(\mu+1)\avk)$.  
In the case of diverging fluctuations the time-scale behavior is therefore
still dominated by $\fluck$ and $\tau$ is always positive whatever
the spreading rate $\lambda$. This allows to recover the absence of an
epidemic threshold, i.e. the lack of a decreasing prevalence region in
the parameter space. 

In order to check these analytical results we have performed numerical
simulations of the SI model on networks generated with the
Barab{\'a}si-Albert (BA) algorithm \cite{barabasi99}.  These networks
have a scale-free degree distribution $P(k)\sim k^{-3}$ and we use
different network sizes $N$ and minimum degree values $m$ in order to
change the level of heterogeneity, that in this case is given by
$\kappa  \sim m \ln N$. Simulations use an agent-based modeling strategy
in which at each time step the SI dynamics is applied to each vertex
by considering the actual state of the vertex and its neighbors.  It
is then possible to measure the evolution of the average number of infected
individuals and other quantities, by averaging over $10^3$ realizations
of the dynamics. In addition, given the stochastic
nature of the model, different initial conditions and networks
realizations can be used to obtain averaged quantities.  In
Fig.~\ref{fig1} we report the early time behavior of outbreaks in
networks with different heterogeneity levels and the behavior of the
measured $\tau$ with respect to $\avk/\lambda(\fluck-\avk)$.  The
numerical results recover the analytical prediction with great
accuracy.  Indeed, the BA network is a good example of heterogeneous
network in which the approximations used in the calculations are
satisfied~\cite{corr}. In networks with correlations we expect to find
different quantitative results but a qualitatively similar framework
as it happens in the case of the epidemic threshold
evaluation~\cite{corr}.

The previous results show that the heterogeneity of scale-free connectivity
patterns favors epidemic spreading not only by suppressing the epidemic
threshold, but also by accelerating the virus propagation in the population.
The velocity of the spreading leaves us with very short response times in the
deployment of control measures and a detailed knowledge of the way epidemics
propagate through the network could be very valuable in the definition of
adaptive strategies. Indeed, the epidemic diffusion is far from homogeneous.
The simple formal integration of Eq.~(\ref{SIk}) written for $s_k=1-i_k$
yields $s_k(t)=s_{k}^0e^{-\lambda k\Phi(t)}$ where $\Phi(t)=\int_0^t d t'
\Theta(t')$.  This result is valid for any value of the degree $k$ and the
function $\Phi(t)$ is positive and monotonically increasing. This last fact
implies that $s_k$ is decreasing monotonically towards zero when time grows.
For any two values $k>k'$, and whatever the initial conditions $s_{k}^0$ and
$s_{k'}^0$ are, there exists a time $t^*$ after which $s_k(t)<s_{k'}(t)$.  This
means that vertices belonging to higher degree classes are generally infected
more quickly.
\begin{figure}[t]
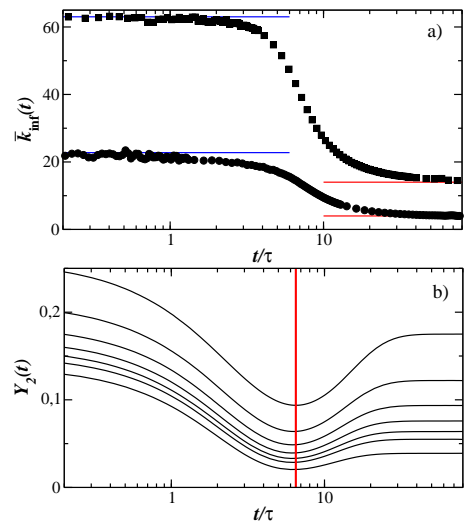

\begin{center}
\epsfig{file=fig2a.eps,width=6cm}
\epsfig{file=fig2b.eps,width=6cm}
\end{center}
\caption{a) Time behavior of the average degree of the newly
  infected nodes for SI outbreaks in BA networks of size $N=10^4$.  Time is
  rescaled by $\tau$. Reference lines are drawn at the asymptotic values
  $\fluck/\avk$ for $t\ll\tau$ and $m$ for $t\gg\tau$. The two curves are for
  $m=4$ (bottom) and $m=14$ (top).  b) Inverse participation ratio $Y_2$
  versus time for BA network of size $N=10^4$ with minimum degree
  $m=4,6,8,10,12,14$ and 20, from top to bottom. Time is rescaled with $\tau$.
  The reference line indicates the minimum of $Y_2$ around $t/\tau\simeq 6.5$.
  }
\label{fig2} 
\end{figure}
A more precise characterization of the epidemic diffusion through the
network can be achieved by studying some convenient quantities in
numerical spreading experiments in BA networks. First, we measure the
average degree of the newly infected nodes at time $t$, defined as
\begin{equation}
\overline{k}_{\rm inf}(t) =\frac{\sum_k k[I_k(t)-I_k(t-1)]}{I(t)-I(t-1)}.
\label{knew}
\end{equation}
In Fig.~\ref{fig2}a) we plot this quantity for BA networks as a
function of the rescaled time $t/\tau$. The curves show an initial
plateau that can be easily understood by considering that at very low
density of infected individuals $i$, each vertex will infect a
fraction of its neighbors without correlations with the spreading from
other vertices. In this case each edge points to a vertex with degree
$k$ with probability $kP(k)/\avk$ and the average degree of newly
infected vertices is given by $ \overline{k}_{\rm inf}(t)=\fluck/\avk$. 
After this initial regime, $\overline{k}_{\rm inf}(t)$ 
decreases smoothly when time increases. The
dynamical spreading process is therefore clear; after the hubs are
very quickly infected the spread is going always towards smaller
values of $k$. This is confirmed by the large time regime that settles
in a plateau $\overline{k}_{ \rm inf}(t)=m$; i.e. the vertices
with the lowest degree are the last to be infected.

Further information on the infection propagation is provided by the
inverse participation ratio $Y_2(t)$ \cite{Derrida:1987}. We first
define the weight of infected individuals in each class degree $k$ by
$w_k(t)=I_k(t)/I(t)$. The quantity $Y_2$ is then defined as
\begin{equation}
  Y_2(t)=\sum_k w_k^2(t).
\label{y2}
\end{equation}
If $Y_2\sim 1/k_{\rm max}$, infected vertices are homogeneously distributed among all
degree classes. In contrast, if $Y_2$ is not small (of order unity) then the
infection is localized on some specific degree classes that dominate the sum
of Eq.~(\ref{y2}). In Fig.~\ref{fig2}b) we report the behavior of $Y_2$
versus time for BA networks with different minimum degree.  The function
$Y_2$ has a maximum at the early time stage, indicating that the infection is
localized on the large $k$ classes, as we infer from the plot of $\langle
k^{\rm inf}(t)\rangle$, see Fig.~\ref{fig2}a).  Afterwards $Y_2$ decreases, with
the infection progressively invading lower degree classes, and providing a
more homogeneous diffusion of infected vertices in the various $k$ classes.
Finally, the last stage of the process corresponds to the capillary invasion
of the lowest degree classes which have a larger number of vertices and thus
provide a larger weight.  In the very large time limit, when the whole
network is infected, $Y_2(t=\infty)=\sum_kP(k)^2$.  Noticeably, curves for
different levels of heterogeneity have the same time profile in the rescaled
variable $t/\tau$.  This implies that, despite the various approximations
used in the calculations, the whole spreading process is dominated by the
time-scale defined in the early exponential regime of the outbreak.

The presented results provide a clear picture of the infection
propagation in heterogeneous networks. First the infection takes
control of the large degree vertices in the network.  Then it rapidly
invades the network via a cascade through progressively smaller degree
classes. The dynamical structure of the
spreading is therefore characterized by a hierarchical cascade from
hubs to intermediate $k$ and finally to small $k$ classes.  This
result, along with the very fast growth rate of epidemic outbreaks,
could be of practical importance in the set-up of dynamic control
strategies in populations with heterogeneous connectivity patterns. In
particular, targeted immunization strategies that evolve with time
might be particularly effective in epidemics control.
\begin{acknowledgments}
  A.B, A. V, and R.P.-S. . are partially funded by the EC-Fet Open
  project COSIN IST-2001-33555. R.P.-S.  acknowledges support from the
  MCyT (Spain), and from the DURSI, Generalitat de Catalunya (Spain).
\end{acknowledgments}

\end{document}